\documentclass[]{spie}  

\usepackage{amsmath,amsfonts,amssymb}
\usepackage{graphicx}
\usepackage{multirow,bigdelim}

\def\CN2{\mbox{$C_N^2 \ $}}

\def\CT2{\mbox{$C_T^2 \ $}}

\def\sigmal2{\mbox{$\sigma ^{2}_{I} \ $}}

\title{New achievements in optical turbulence forecast systems in operational mode}

\author[a]{Elena Masciadri}
\author[a]{Alessio Turchi}
\author[a]{Gianluca Martelloni}

\affil[a]{INAF - Osservatorio Astrofisico di Arcetri, L.go E. Fermi 5, 50125  Florence, Italy}

\authorinfo{Correspondence to Elena Masciadri: elena.masciadri@inaf.it}

\begin{document} 
\maketitle 

\begin{abstract}
In this contribution, we present the most recent progresses we obtained in the context of a long-term program we undertook since a few years towards the implementation of operational forecast systems (a) on top-class ground-based telescopes assisted by AO systems to support the flexible scheduling of observational scientific programs in night as well in day time and (b) on ground-stations to support free space optical communication. Two topics have been treated and presented in the Conference AO4ELT6:\newline

1. ALTA is an operational forecast system for the OT and all the critical atmospheric parameters affecting the astronomical ground-based observations conceived for the LBT. It operates since 2016 and it is in continuous evolution to match with necessities/requirements of instruments assisted by AO of the LBT (SOUL, SHARK-NIR, SHARK-VIS, LINC-NIRVANA,...). In this contribution, we present a new implemented version of ALTA that, thanks to an auto-regression method making use of numerical forecasts and real-time OT measurements taken in situ, can obtain model performances (for forecasts of atmospherical and astroclimatic parameters) never achieved before on time scales of the order of a few hours. \newline

2. We will go through the main differences between optical turbulence forecast performed with mesoscale and general circulation models (GCM) by clarifying some fundamental concepts and by correcting some erroneous information circulating recently in the literature. \newline

\end{abstract}



\section{INTRODUCTION}
\label{sec:intro} 

The goal of this contribution is to deal about the progresses achieved recently in the field of the optical turbulence forecast applied to the ground-based astronomy by our team. We focused our attention on a couple of arguments indicated in the Abstract. As in the last months after the conference, the topic 1 has been extensively treated in a paper submitted to a peer-reviewed journal we mention here just the main elements related to the preliminary results presented at the Conference and we refer the readers to the peer-reviewed journal paper for the complete analysis.

In the second part of the contribution, we go through the main differences between forecasts provided by GCMs and mesoscale models trying to clarify some fundamental concepts related to the two tools and trying to correct some misleading discussions circulating in the literature. 

\section{New achievements: operational forecasts at different time scales}

\begin{figure} [ht]
\begin{center}
\begin{tabular}{cc} 
\includegraphics[width=10cm]{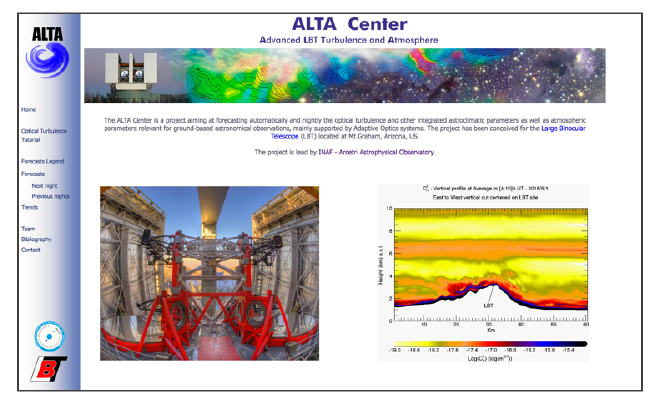}\\
\end{tabular}
\end{center}
\caption
{\label{fig:temp_surf} ALTA Center home page - http://alta.arcetri.astro.it.  }
\end{figure} 

\begin{figure} [ht]
\begin{center}
\begin{tabular}{cc} 
\includegraphics[width=16cm]{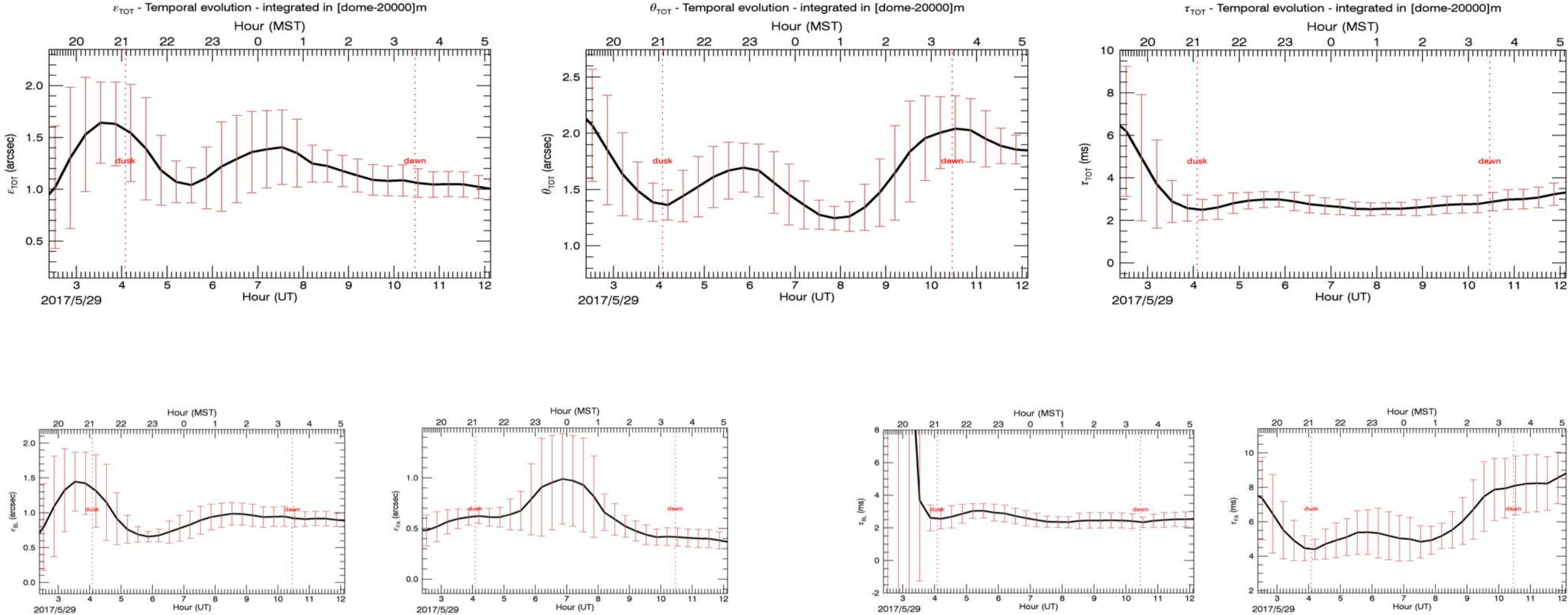}\\
\end{tabular}
\end{center}
\caption
{\label{fig:astroclim} Temporal evolution during a night of the three most relevant astroclimatic parameters: the seeing $\varepsilon$ integrated on the whole atmosphere (top-left), the isoplanatic angle $\theta_{0}$ (top-center) and the wavefront coherence time $\tau_{0}$ integrated on the whole atmosphere (top-right). Temporal evolution of the seeing integrated in the boundary layer and in the free atmosphere (bottom left). Temporal evolution of the $\tau_{0}$ in the boundary layer and in the free atmosphere (bottom right). The seeing and the isoplanatic angle are expressed in arcsec; the wavefront coherence time in msec. Raw temporal frequency is two minutes; data are re-interpolated on a time scale of 20 minutes; the error bar indicate the standard deviation calculated on the raw data set. Temporal evolution is shown between the sunset and the sunrise, red dashed lines indicate the dusk and the dawn.}
\end{figure} 

\begin{figure} [ht]
\begin{center}
\begin{tabular}{cc} 
\includegraphics[width=13cm]{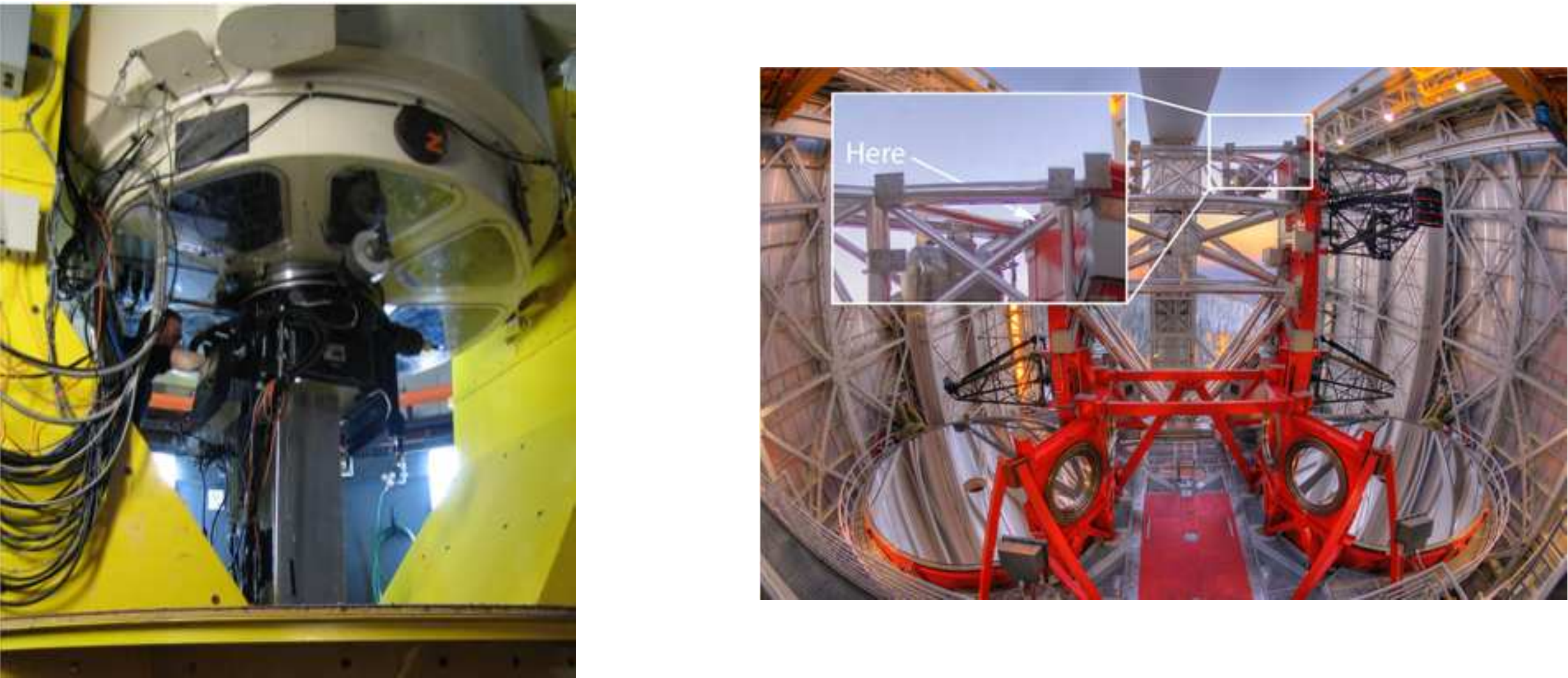}\\
\end{tabular}
\end{center}
\caption
{\label{memo} Left: Generalized SCIDAR used to monitor the $\CN2$ during 43 nights in the period 2005-2008 at Mt.Graham. The instrument is located at the focus of the VATT telescope located on the summit of Mt.Graham at around 200~m from the LBT. Right: DIMM running nightly at Mt.Graham measuring the seeing. The instrument is located on the top of the LBT dome (see zoom in the square with white frame.) }
\end{figure} 

ALTA Center is an automatic and operational forecast system conceived to support the Large Binocular Telescope (LBT), located at Mt.Graham, Arizona, US (Fig.\ref{fig:temp_surf}). ALTA Center performs forecast of the optical turbulence and the atmospherical parameters relevant for the ground-based astronomy. An example of products that can be found in the web-site is shown in Fig.\ref{fig:astroclim}. The automatic and operational system is conceived to provide the forecast early in the afternoon for the next night. The calibration and the validation of the Astro-Meso-Nh model has been performed using, as a reference, observations from a Generalized SCIDAR and a DIMM (Fig.\ref{memo}). The first instrument requires a telescope of at least 1~m in diameter and it has been indeed run at the focus of the Vatical Advanced Technology Telescope for 43 nights uniformly distributed in different periods of the year between 2005 and 2008. The DIMM is a monitor that runs nightly at LBT.

The study presented in this contribution is triggered by the consideration that the most critical time scale on which it should be useful to know in advance the turbulence conditions for the implementation of the flexile-scheduling for all the applications involving ground-based astronomy supported by adaptive optics is of 1-2 hours. The goal of this study aims to answer to the following question: is it possible to provide forecast at a time scale of 1h-2h and possibly improving the model performances on this time scale ?  We investigated the impact obtained on model performances by using the autoregressive method that belongs to the filtering techniques such as Kalman filter and machine learning approaches. This techniques takes advantage of a simultaneous presence of a forecast obtained with the numerical model and the real-time measurements provided by in-situ instrumentation. In our case, we considered the forecast obtained with the Astro-Meso-Nh model in the standard configuration. We mean with that a forecast produced early in the afternoon for the whole successive night as described, for example, in Masciadri et al. 2020 [\cite{masciadri2019}].

The content of this analysis has been synthetised on a paper submitted to a peer-reviewed journal\cite{masciadri2019}. We refer the reader to this paper.

\section{Forecasts with mesoscale models and general circulation models} 

\begin{figure} [ht]
\begin{center}
\begin{tabular}{cc} 
\includegraphics[width=7cm,angle=-90]{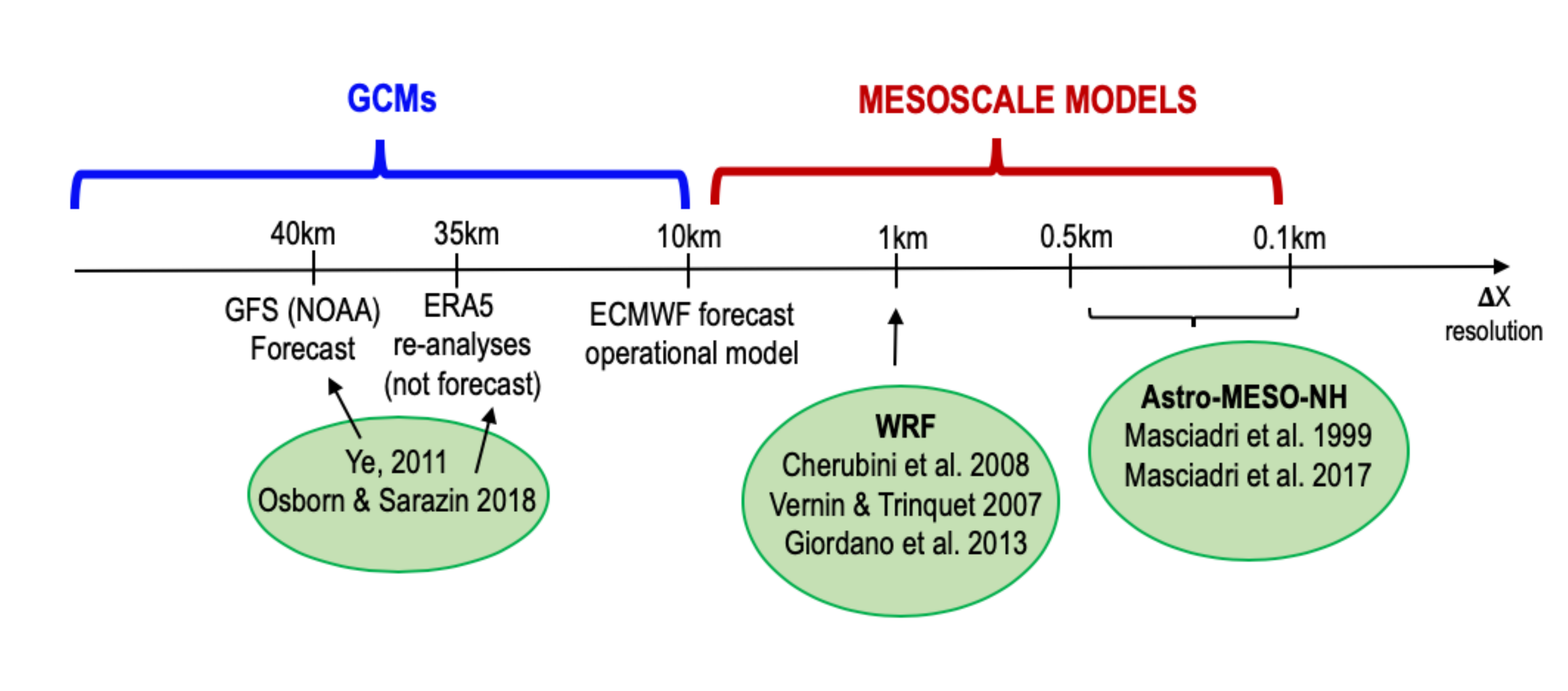}\\

\end{tabular}
\end{center}
\caption
{\label{fig:models} Synthetic scheme showing the different typology of models used to forecast the optical turbulence
}
\end{figure} 

There are two elements that plays a fundamental role in the optical turbulence forecast: from one side the typology of numerical models used to reconstruct the vertical or volumetric distribution of the $\CN2$ and, on the other side, the physics and the algorithms used to described numerically the optical turbulence, more precisely the $\CN2$. In Fig.\ref{fig:models} is shown a synthetic panorama referring to a list of papers dealing with different approaches in the field of the turbulence forecasts in the astronomical context. 

The two main categories of models are those of the General Circulation Models (GCMs) and the mesoscale models\footnote{We skip-off in this context the Large Eddy Simulations (LES) that represents a category of models for which the turbulence might be partially resolved and partially parameterized}. GCMs are applied to the whole Earth and they have a horizontal resolution of the order of 10-20 km or more (at present the model with the highest horizontal resolution is the HRES of the European Centre for Medium Range Weather Forecasts (ECMWF) with a resolution of 9-10 km depending on the latitudes). Mesoscale models are applied to limited regions of the Earth, they are frequently used in grid-nesting configuration i.e. a set of imbricated domains in which each successive internal domain has a smaller extension and a higher resolution with respect to the precedent domain. Such a solution aims to increase the resolution in a region close to the point of interest but at the same time to maintain reasonable calculation time. Mesoscale model have been conceived and invented exactly to overcome evident limitations of the GCMs. They are commonly used in the meteorological context and it is more and more frequent their use in an operational approach. The more evident advantage of this kind of models is the horizontal resolution. Mesoscale models can achieve resolutions much higher than GCMs. 

Even if the highest resolution in the traditional classification is of the order of 1-2~km, in the most recent years, we can easily find sub-kilometric resolutions as high as 100~m therefore a factor 100 higher with respect to the GCMs. This is due to the fact that GCMs resolution increases too and, as a consequence, it is possible to conceive more aggressive configurations with mesoscale models. The higher resolution of mesoscale models permits to definitely better reconstruct the physics of the atmospheric flow, particularly that developing in the boundary layer and, in general, in the low part of the atmosphere with respect to the GCMs. The topography is much better correlated to the reality and, especially on mountain regions (as is the case in many astronomical sites), this makes a great difference. A higher resolution permits to reconstruct the gravity and mountain waves as well as the strong wind speed shear close to the ground in proximity of isolated summits in a much more realistic way. We know that astronomical sites are frequently placed on isolated peaks. All these phenomena are not well resolved by GCMs that inevitably smooth out the topography and affects in a non negligible way phenomena developing in the surface and boundary layer. 

The second important element that characterises the forecast of the optical turbulence is the algorithm or method used to forecast the OT. This 'can be' totally independent on the category of model used. The two main approaches used in the astronomical context are: (1) the prediction of the optical turbulence is based on the prognostic turbulent kinetic equation (TKE), on the definition of a mixing length and on the parameterization of the OT. This is the approach widely treated by Masciadri et al.[ \cite{masciadri1999,masciadri2001,masciadri2017}] and Cherubini et al. [\cite{cherubini2008}] with some differences using different mesoscale models; (2) the OT is expressed as a function of the macroscopic temperature and wind speed using empirical laws. This is the approach followed by Vernin \& Trinquet[\cite{trinquet2007}] and Giordano et al.[\cite{giordano2013}] with mesoscale models and by Ye[\cite{ye2011}] and Osborn \& Sarazin[\cite{osborn2018}] with GCMs.

Goals of this contribution is to try to point out some erroneous statements circulating in the literature or in the astronomical community. The fact that the numerical weather prediction is obviously not a common discipline specific of the astronomical context, some misunderstandings are physiologic. We will go through rapidly on some macroscopic 'false believes'.\\

\begin{figure} [ht]
\begin{center}
\begin{tabular}{cc} 
\includegraphics[width=13cm]{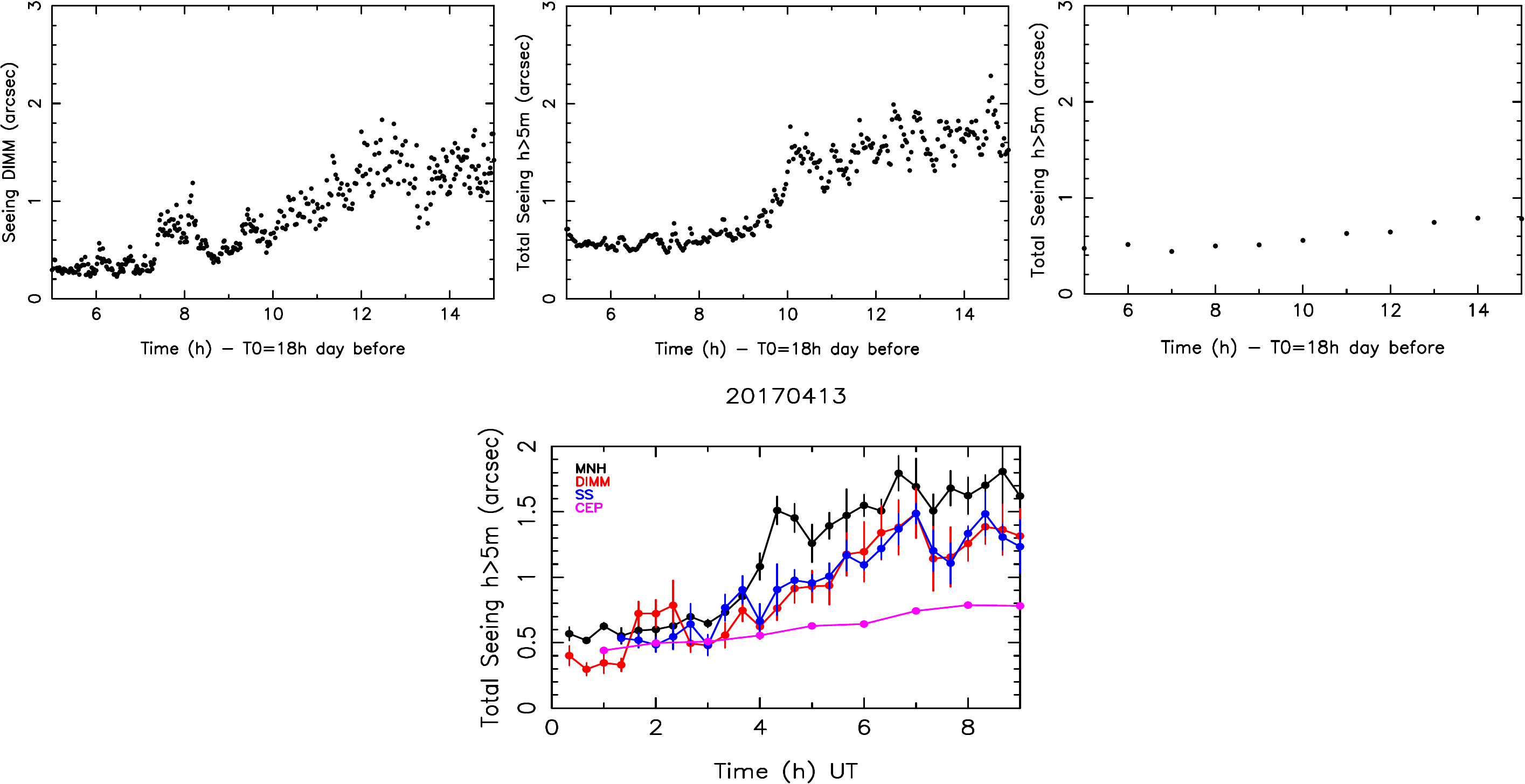}\\
\end{tabular}
\end{center}
\caption
{\label{fig:dimm_mnh_cep_20170413} Top-left: temporal evolution of the seeing measurements obtained with a DIMM during the night 20170413; top-centre: temporal evolution of the forecast calculated by the Astro-Meso-Nh. Temporal frequency of the forecast is 2 minutes; top-right: forecast calculated by the ECMWF General Circulation Model (GCM). Temporal frequency of the forecast is 1h. Bottom: temporal evolution of measurements and forecasts obtained with different instruments and models. Observations and forecasts have been treated with a moving average of 1 h and a resampling of 20 minutes.}
\end{figure} 

\begin{figure} [ht]
\begin{center}
\begin{tabular}{cc} 
\includegraphics[width=13cm]{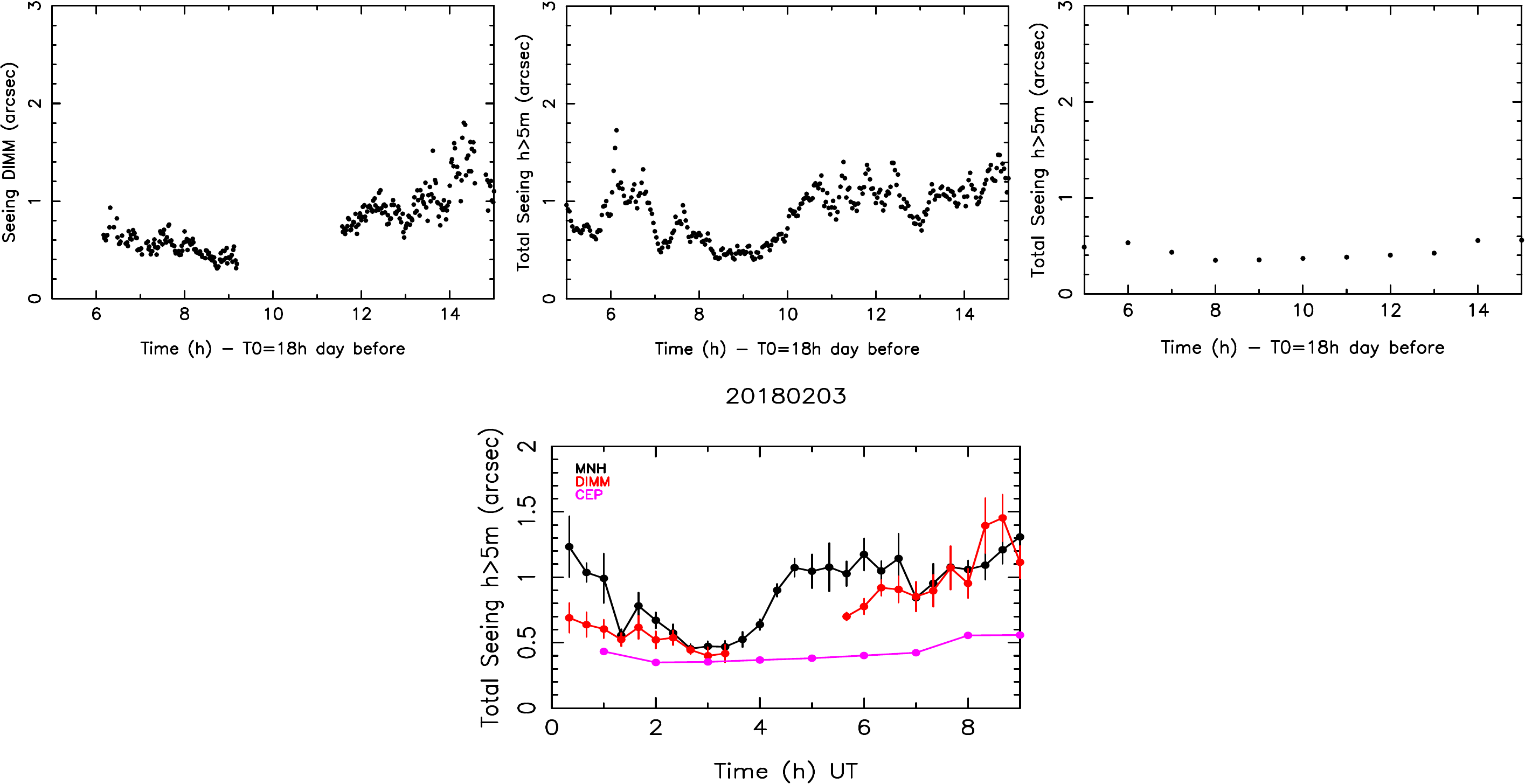}\\
\end{tabular}
\end{center}
\caption
{\label{fig:dimm_mnh_cep_20180203} Same as Fig.\ref{fig:dimm_mnh_cep_20170413} but for the night 20180203}
\end{figure} 

\begin{figure} [ht]
\begin{center}
\begin{tabular}{cc} 
\includegraphics[width=13cm]{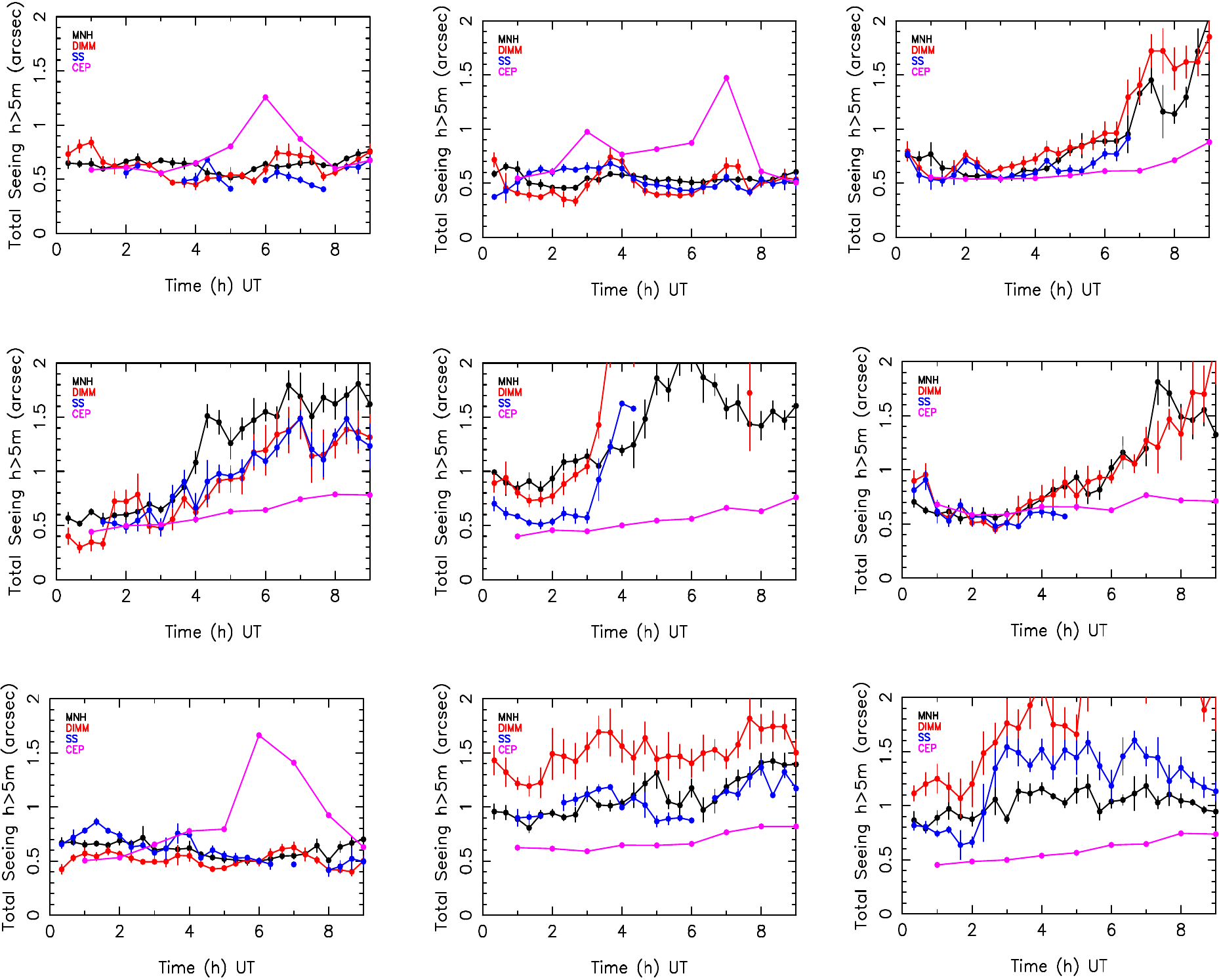}\\
\end{tabular}
\end{center}
\caption
{\label{fig:serie_nights} Temporal evolution of the total seeing as measured by the DIMM and the Stereo-SCIDAR and the seeing forecast by the Astro-Meso-Nh and the GCM of the ECMWF for nine different nights. For the Astro-Meso-Nh the forecast frequency is of 2 minutes. Data are resampled with 20 minutes and the error bar indicate the variability of the model. For the ECMWF GCM, we considered the sequence of forecasts with the shortest delay time (6h) and the frequency of 1h.  }
\end{figure} 

\begin{figure} [ht]
\begin{center}
\begin{tabular}{cc} 
\includegraphics[width=5cm]{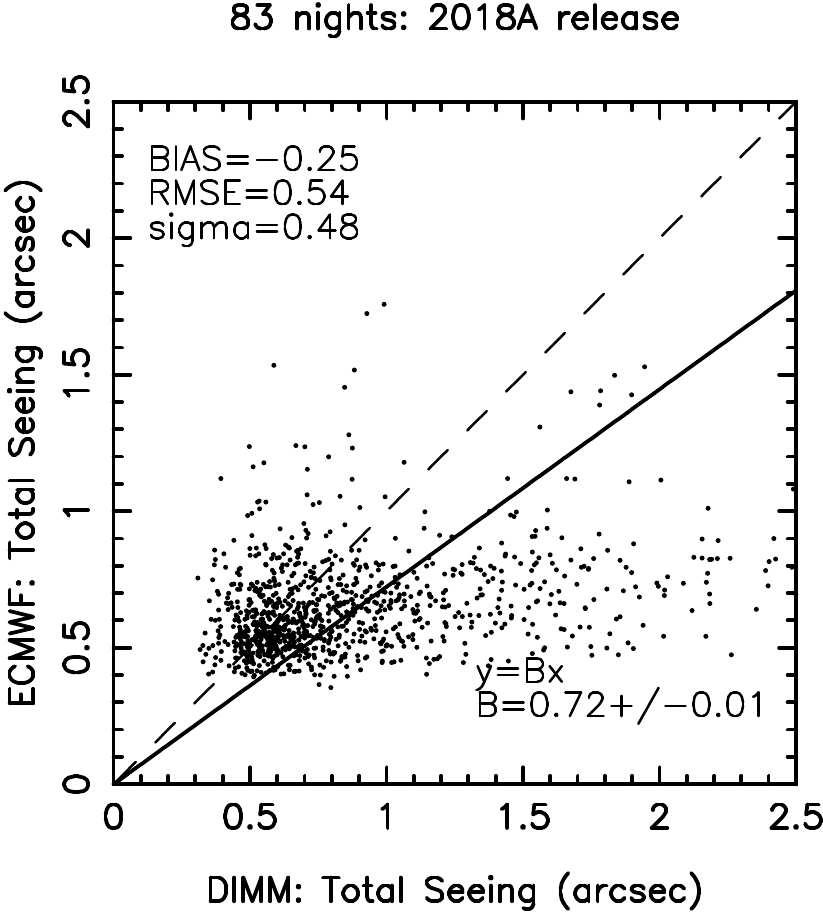}\\%
\end{tabular}
\end{center}
\caption
{\label{fig:cep} Scattering plot of seeing measured by a DIMM and forecasted with a GCM on a sample of 83 nights}
\end{figure} 

(A) Concerning the advantage or disadvantages of a category of a models with respect to the others here an example that we hope can clarify things. It has been affirmed[\cite{osborn2018}] 
that GCMs better put in evidence rapid change of the atmosphere because mesoscale model requires supplementary calculation. That statement risks to be very misleadings for a community not used to deal about this topic. It is true that the use of a mesoscale model implies an additional calculation but we remind that the mesoscale models have been invented exactly to improve results obtained with GCM. The operational weather forecast at national level is done today with mesoscale models (for example AROME in France). That indicates clearly that the additional computation time is visibly not a major problem and it is hard to belief that GCM better put in evidence rapid changes of the atmosphere than mesoscale models. It should be true the opposite. We remind, indeed, that the critical element in the OT forecast is not the calculation time (provided it is within reasonable values) but the temporal frequency of the forecast and the global accuracy of the system. It is perfectly possible that a prediction requiring a longer calculation time is more accurate than one that requires a shorter time scale.  One can have a prediction of X consecutive hours performed in $\Delta$T1 time before a reference time T$_{X}$ that is more accurate than a prediction of X hours calculated in $\Delta$T2 time before T$_{X}$ with ($\Delta$T2 $<$ $\Delta$T1). The following example can clarify the concept. 

In Fig.\ref{fig:dimm_mnh_cep_20170413}-left we report, as an example, the temporal evolution of the seeing as measured during the night 13/4/2017 with a DIMM at Cerro Paranal, the site of the VLT. The temporal frequency of the measurement is typically of the order of 1 minutes. In Fig.\ref{fig:dimm_mnh_cep_20170413}-center is shown the temporal evolution of the same night as predicted by the Astro-Meso-Nh using the method[\cite{masciadri2017}] (hereafter M17) in the configuration used for the operational system of ALTA Center (resolution of 0.5~km). The forecast of the next night is accessible early in the afternoon (at around 14:00 LT). The temporal frequency is 2 minutes (but it is easily possible to increase the frequency is requested). In Fig.\ref{fig:dimm_mnh_cep_20170413}-right is shown the temporal evolution of the same night as predicted by the GCM of the ECMWF (resolution of 10~km) using, for example, the equation presented in[\cite{osborn2018}] (hereafter OS18) considering the shortest sequence of predictions on a time scale of 6h. The frequency of outputs is 1 hour. We highlight that we use here real forecast, not re-analyses as in the case of OS18. The same exercise is repeated in Fig.\ref{fig:dimm_mnh_cep_20180203} for the night 3/2/2018.

As can be seen in Fig.\ref{fig:dimm_mnh_cep_20170413} and Fig.\ref{fig:dimm_mnh_cep_20180203} the Astro-Meso-Nh model is much better correlated to observations and definitely better reconstructs the temporal trend of the OT than the GCM (OS18). The Astro-Meso-Nh approachs reacts in a much better way to a rapid changes of the atmosphere conditions than the GCM (with OS18) approach. This is because the mesoscale model better reconstructs the atmosphere (particularly the low part of the atmosphere) with respect to GCMs. Besides that, a frequency of 2 minutes permits to reconstruct a rapid change of the atmosphere in a much faster way that a frequency of 1 hour. Fig.\ref{fig:serie_nights} shows the temporal evolution of the measurements by the DIMM and the Stereo-SCIDAR and the seeing forecast by the Astro-Meso-Nh and the GCM of the ECMWF for nine different nights. This figure clearly shows that what indicated in Fig.\ref{fig:dimm_mnh_cep_20170413} and Fig.\ref{fig:dimm_mnh_cep_20180203} are not isolated cases.  The total seeing forecasts obtained with the Astro-Meso-Nh model are much better correlated to those obtained with the GCM of the ECMWF. Also in those cases in which some discrepancies between measurements can be observed (see last row in Fig.\ref{fig:serie_nights}, center and right with red and blue lines), the Astro-Meso-Nh shows a better reaction to rapid changes of the atmosphere and a better correlation to measurements than ECMWF GCM. 
This results is not surprising and proves that is deeply misleading to claim that GCMs are preferable to mesoscale ones because they permit a minimal calculation time. 
In Fig.\ref{fig:cep} we report the scattering plot of predictions versus measurements on the whole sample of measurements (DIMM) related to the site testing campaign that took place at Cerro Paranal (Release 2018A)[\cite{osborn2018}]. As can be seen, the RMSE is 0.54" i.e. pretty large. The important thing to retain is that, as soon as there is a variability, the level of decorrelation is very high. We observe two halos with observations having very high values associated to weak model values and viceversa.

We highlight the fact that while the predictions provided by Astro-Meso-Nh (with M17) shown in Fig.\ref{fig:dimm_mnh_cep_20170413}, to Fig.\ref{fig:serie_nights} (black lines) are displayed at 14:00 LT (18:00 UT) i.e. early in the afternoon, the forecast shown in the same figures obtained with the GCMs (OS18) (pink lines) is a composition of outputs that are accessible at different times. More precisely, the pink line is a composition of forecast available at different time scales. Forecasts up to 00:00 UT (20:00 LT) are available at 14:00 LT;  forecasts between 01:00 UT and 06:00 UT are available at 21:00 LT; forecasts between 07:00 UT and the end of the night are available at 03:00 LT. We have therefore three regimes associated to three different local time: 14:00 LT, 21:00 LT and 03:00 LT. 
We report this example just to prove that it is extremely delicate to do "fair" comparisons between different solutions. Here we see that, even with a larger delay time, mesoscale model forecast are more accurate than GCMs forecasts.

(B) It has been said in the literature[\cite{osborn2018}] that the Astro-Meso-Nh model[\cite{masciadri2017}] requires 15h to reach the thermodynamic equilibrium. In reality the calculation time is much shorter than 15h and in any case the calculation time has obviously nothing to see with the thermodynamic equilibrium. Besides that, as we showed in the examples just discussed at the bullet (A), the duration of the calculation time is not, by itself a negative element. The important thing is the final performance of the model. 

(C) Some confusion has been done on the concept of calibration[\cite{osborn2018}]. The model calibration has been introduced in the past to optimise the configuration of a model for specific sites. The fact that a model is calibrated does not mean that it is not possible to use an universal calibration. A necessary condition to be able to talk about a universal calibration is to have rich statistical samples of measurements (vertical profilers) on different sites in the world and with different instruments that are not accessible at present. This is the reason why, the problem of the universal calibration has been simply skipped-off so far and the forecast approach has been tackled so far by calibrating models to try to do the best for each site using the available measurements. Many different formulations of calibration has been proposed during the years. The day in which it will be possible to collect enough heterogeneous data on different sites we will be able to verify if it is possible to use a  universal calibrations. In other words, availability of data is just a necessary but not sufficient condition to guarantee an universal calibration. So far 'all' the methods presented in the literature, no one excluded, at least in the astronomical context, use a specific calibration. 

(D) Another false belief circulating in the community is that a GCM can be used without the support of measurements while mesoscale models need a validation for each site. That is again a no sense.  The fact that a model is extended on the whole Earth changes nothing. A mesoscale model can be applied to whatever place in the world therefore the real question is if a model has been validated to forecast a specific parameter or not and with which level of performances. A model can be used as a reference if the model has been proven to be reliable, independently on the fact that the model is extended on the whole Earth or on a limited area model.

\section{CONCLUSIONS}

In this contribution we presented preliminary results on a new method to forecast astroclimatic parameters and meteorologic parameters at short time scales (1h-2h). Extended analysis and complete results will appear soon on a peer-reviewed journal paper. In the contribution we report the synthesis of the second topic presented in the talk. More precisely we tried to clarify some fundamental concepts related to the atmospherical modelling  performed with different typologies of models and to correct some conceptual wrong believes circulating in the astronomical community. 

\acknowledgments       
We thank the ``Large Binocular Telescope Corporation'' (http://www.lbto.org/) for supporting and financing the ALTA project. \\


\end{document}